\documentclass[12pt]{article}

\usepackage[margin=2cm]{geometry}
\usepackage[latin1]{inputenc}
\usepackage{graphicx}
\usepackage{fancyhdr}
\usepackage{amsfonts}
\usepackage{apacite}
\usepackage{color} 

\newcommand{\bfx}{\mathbf{x}}

\newcommand{\bff}{\mathbf{f}}

\newcommand{\bfF}{\mathbf{F}}

\newcommand{\bfA}{\mathbf{A}}

\newcommand{\bfTheta}{\mathbf{\Theta}}

\definecolor{maroon}{rgb}{0.6,0,0}

\author{
Ueli Rutishauser\\
Department of Neural Systems, Max Planck Institute for Brain Research\\ 
\texttt{urut@brain.mpg.de} \\
\and
Jean-Jacques Slotine \\
Nonlinear Systems Laboratory, Massachusetts Institute of Technology\\
\texttt{jjs@mit.edu} \\
\and
Rodney J. Douglas\\
Institute of Neuroinformatics, ETH Zurich and University of Zurich\\
\texttt{rjd@ini.phys.ethz.ch}\\
}

\title{Competition through selective inhibitory synchrony\footnote{This article has been accepted by MIT Press for publication in a future issue of Neural Computation (2012). This is a pre-print version (as accepted). This version was updated on 04/03/12, correcting minor typos discovered during proof-reading.}}

\begin{document}
\maketitle

\section{Abstract}
Models of cortical neuronal circuits commonly depend on inhibitory feedback to control gain, provide signal normalization, and to selectively amplify signals using winner-take-all (WTA) dynamics. Such models generally assume that excitatory and inhibitory neurons are able to interact easily, because their axons and dendrites are co-localized in the same small volume. However, quantitative neuroanatomical studies of the dimensions of axonal and dendritic trees of neurons in the neocortex show that this co-localization assumption is not valid. In this paper we describe a simple modification to the WTA circuit design that permits the effects of distributed inhibitory neurons to be coupled through synchronization, and so allows a single WTA to be distributed widely in cortical space, well beyond the arborization of any single inhibitory neuron, and even across different cortical areas. We prove by non-linear contraction analysis, and demonstrate by simulation that distributed WTA sub-systems combined by such inhibitory synchrony are inherently stable. We show analytically that synchronization is substantially faster than winner selection. This circuit mechanism allows networks of independent WTAs to fully or partially compete with other.

\section{Introduction}
Many models of neuronal computation involve the interaction of a population of excitatory neurons whose outputs drive inhibitory neuron(s), which in turn provide global negative feedback to the excitatory pool \cite{Amari77b, Douglas95, Hahnloser00, YuilleGeiger03, Maass00, Hertz91, Rabinovich00, RutishauserDouglas2010, Coultrip92}. Practical implementation of such circuits in biological neural circuits depend on co-localization of the excitatory and inhibitory neurons, an assumption which studies of the extents of axonal and dendritic trees of neurons in the neocortex show is not valid \cite{KatzelMiesenbock11,Binzegger04,Shepherd05,Douglas2004_neuronal}. Firstly, a substantial fraction of the axonal arborization of a typical excitatory 'spiny' pyramidal neuron extends well beyond the range of the arborization of a typical 'smooth' inhibitory neuron, particularly in the populous superficial layers of the neocortex \cite{YabutaCallaway98,Binzegger04}. This spatial arrangement means that excitatory effects can propagate well outside the range of the negative feedback provided by a single inhibitory neuron. Secondly, the horizontally disposed 'basket' type of inhibitory neuron, which is a prime candidate for performing normalization, makes multiple synaptic contacts with its excitatory targets, so that even within the range of its axonal arborization, not all the members of an excitatory population can be covered by its effect.  This connection pattern means that excitatory neurons within some local population must either be partitioned functionally, or multiple smooth cells must co-operate to cover the entire population of excitatory cells.

In previous publications we have shown how winner-take-all (WTA) circuits composed of a small population of excitatory neurons and a single inhibitory neuron can be combined to construct super-circuits that exhibit finite state-machine (FSM) like behavior \cite{RutishauserDouglas2009,Neftci_etal10}. The super-circuits made use of sparse excitatory cross-connections between WTA modules to express the required states of the FSM.  These excitatory connections can extend well outside of the range of the local WTA connections, and so are consistent with the observed long-range lateral excitatory connections referred to above. On the other hand, we have not previously confronted the question of whether the WTA is necessarily localized to the extent of the smooth-cell arborization, or whether the WTA can itself be well distributed in space within or between cortical lamina, or even between cortical areas. In this paper we describe a simple modification to the WTA circuit design that couples the effects of distributed inhibitory neurons through synchronization, and so permits a WTA to be widely distributed in cortical space, well beyond the range of the axonal arborization of any single inhibitory neuron, and even across cortical areas. We also demonstrate that such a distributed WTA is inherently stable in its operation.

\section{Results}

\begin{figure}
\centering
\includegraphics[angle=0,width=18cm]{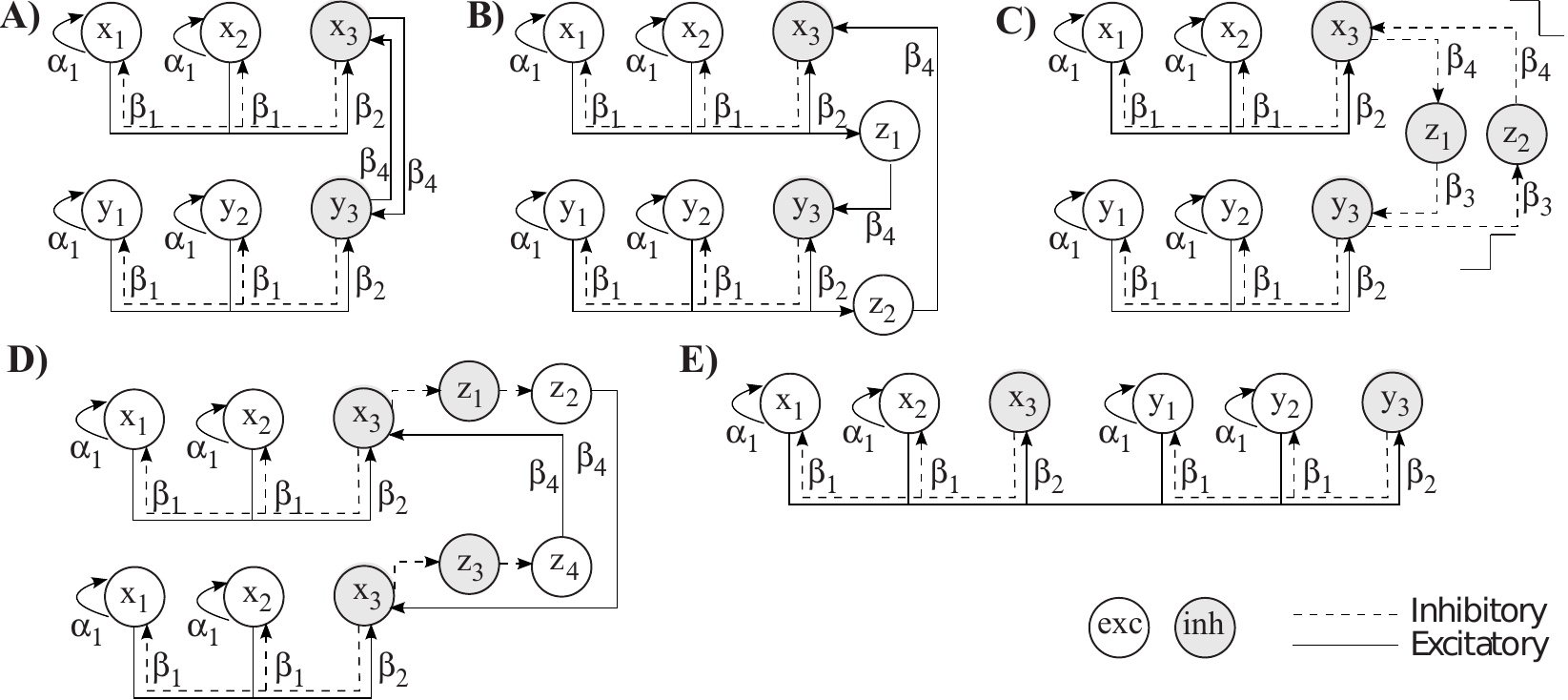}
\caption{Circuits for distributing WTAs. (A) Illustration of the principal idea - mutual excitation of the inhibitory neurons. 
(B-E) Are are biologically plausible versions.
(B) Implementation using intermediate excitatory neurons $z_{1,2}$. This circuit will be considered in detail in Fig. \ref{fig:ElegantCircuit}B with more realistic connectivity.
(C) Implementation using disinhibition of persistently active units $z_{1,2}$ as illustrated by the step-functions.
(D) Implementation with disinhibition and long-range excitatory units.
(E) Implementation using multiplication of inhibitory neurons. Here, $x_3=y_3$ at all times. The maximal excitatory projection length is double that of the inhibitory. 
}
\label{fig:AllCircuits}
\end{figure}

\begin{figure}
\centering
\includegraphics[angle=0,width=15cm]{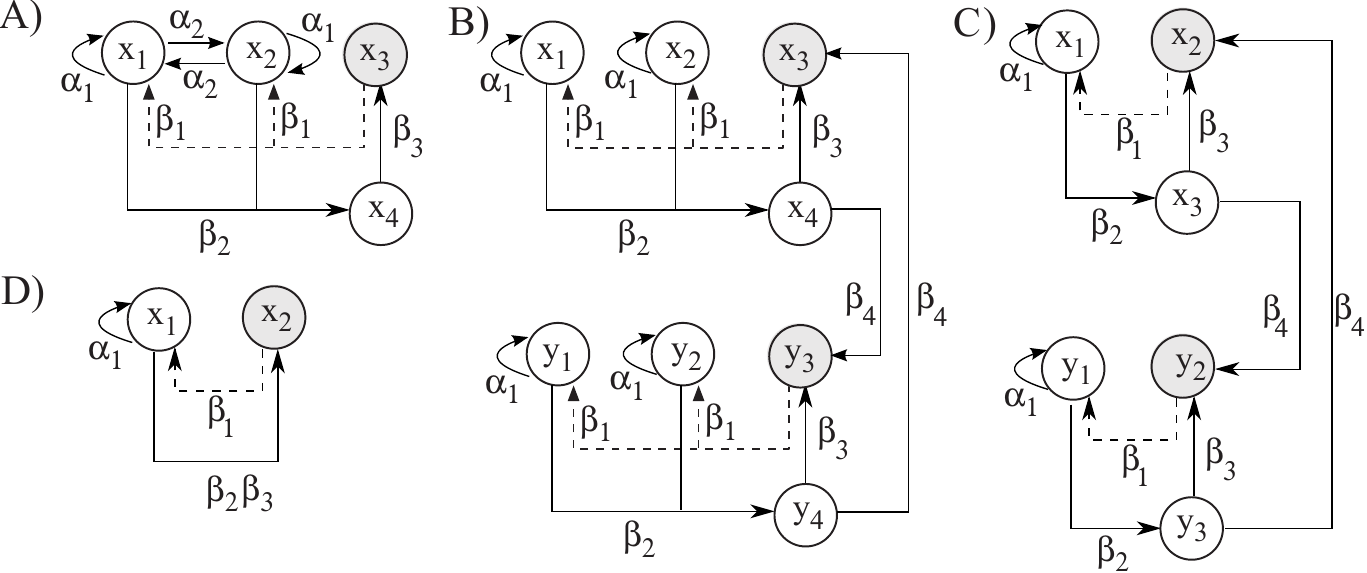}
\caption{Schematic of connectivity. Gray units and dashed lines are inhibitory, white units and straight lines are excitatory. 
(A) Single WTA with two possible winners $x_{1,2}$, inhibitory unit $x_3$, and intermediate excitatory unit $x_4$ that carries the average
activity of $x_{1,2}$. 
(B) Coupling of two WTAs to form a single WTA with 4 possible winners. $\beta_4$ are excitatory long-range connections that
serve to synchronize the two inhibitory units. (C-D) Are reduced versions for theoretical analysis.
(C) Merged WTA with one winner each and thus two possible winners $x_1$ and $y_1$ in the merged network. 
(D) Reduced single WTA.    }
\label{fig:ElegantCircuit}
\end{figure}

We have considered a number of circuits that could be used to distribute spatially the WTA behavior (Fig \ref{fig:AllCircuits}). However, we will describe and analyze only the circuit shown in Fig. \ref{fig:ElegantCircuit}, which we consider to be the most elegant of the distributive mechanisms (notice the similarity to Fig. \ref{fig:AllCircuits}B). The key insight is the following: Under normal operating conditions, all the participating distributed inhibitory neurons should receive the same summed excitatory input. We achieve this by interposing an excitatory neuron in the negative feedback loop from the excitatory population to its local inhibitory neuron. Instead of the local inhibitory neuron summing over its excitatory population, the interposed neuron performs the summing and passes its result to the inhibitory neuron. This result is also copied to the more distant inhibitory neurons in the spatially distributed WTA. In this way the inhibitory neuron of each sub-WTA sums over the projections from the interposed excitatory neurons of all other sub-WTAs, including its own one.  Thus, each inhibitory neuron is able to provide feedback inhibition to its local sub-WTA that is proportional to the total excitation provided by all excitatory neurons participating in the entire distributed WTA. We will show that functionally this amounts to a form of synchrony between all the inhibitory units.

\subsection{Connectivity and dynamics - single WTA}
All the circuits of Fig \ref{fig:AllCircuits} can achieve a distributed WTA by merging several independent WTAs, but we consider the circuit shown in Fig \ref{fig:ElegantCircuit}B to be most feasible, and so our analysis will focus on this one. However, similar reasoning could be applied to all the variants shown. Note that our chosen circuit is similar to that of Fig \ref{fig:AllCircuits}B, but has a more realistic connectivity pattern in that the summed excitatory activity is only projected onto a single unit, which requires less wiring specificity than Fig \ref{fig:AllCircuits}B.

The dynamics of a single WTA (Fig \ref{fig:ElegantCircuit}A) with in total $N$ units, consisting of $1..N-2$ excitatory units, one inhibitory unit $x_{N-1}$ and one intermediary interconnect excitatory unit $x_{N}$,  are

\begin{eqnarray}
\tau \dot{x}_i + G x_i = f( I_i + \alpha x_i - \beta_1 x_{N-1} - T_i) \nonumber \\
\tau \dot{x}_{N-1} + G x_{N-1}  = f( \beta_3 x_{N} - T_{N-1}) \nonumber \\
\tau \dot{x}_{N} + G x_{N}  = f( \beta_2\sum_{j=1}^{N-2} x_j - T_N)
\label{eq:recmapEqs}
\end{eqnarray}

Each excitatory unit receives recurrent input from itself ($\alpha_1$) and its neighbors ($\alpha_{2,3,...}$, see Fig \ref{fig:ElegantCircuit}A). For simplicity, only self-recurrence is considered here ($\alpha=\alpha_1$ and $\alpha_{2,3,...}=0$), but very similar arguments obtain when recurrence from neighboring units is included. 
\noindent Using the weight matrix $\mathbf{W}$ the dynamics of this system is described as
\begin{equation}
\tau \dot{\mathbf{x}} + G \mathbf{x}  = f( \mathbf{W}\mathbf{x} - \mathbf{T} + \mathbf{I}(t) )
\label{eq:dynSingle}
\end{equation}
\noindent with
\begin{equation}
\mathbf{W} = \left[ 
\begin{array}{cccc}
\alpha   & 0       & -\beta_1 & 0 \\
0        & \alpha  & -\beta_1 & 0 \\
0        & 0       & 0        & \beta_3 \\
\beta_2  & \beta_2 & 0        & 0 
\end{array} 
\right]
\label{eq:Wsingle}
\end{equation}
\noindent where the order of units is $x_{1,2,3,4}$ (i.e. first column and row is $x_1$, last column/row is $x_4$).
The firing rate activation function $f(x)$ is a non-saturating rectification non-linearity $max(0,x)$.
We assume $\tau=1$ and $G=1$ throughout unless mentioned otherwise. $\mathbf{T}=[T_1,...,T_{N-1},T_N]$ is a vector of the constant activation thresholds $T_i \geq 0$.

\subsection{Connectivity and dynamics - coupled WTA}

Two identical single WTAs, each described by weight matrices $\mathbf{W}_1=\mathbf{W}_2=\mathbf{W}$ can be combined into one distributed WTA that acts functionally as a single WTA by adding a recurrent excitatory feedback loop $\beta_4$ between the two WTAs (Fig \ref{fig:ElegantCircuit}B). The weight matrix $\mathbf{W}_C$ of the merged system is
\begin{equation}
\mathbf{W}_C = \left[ 
\begin{array}{cc}
\mathbf{W}_1 & \mathbf{C}_2 \\ 
\mathbf{C}_1 & \mathbf{W}_2
\end{array} 
\right]
\label{eq:Wcoupled1}
\end{equation}
\noindent with interconnections
\begin{equation}
\mathbf{C}_1 = \left[ 
\begin{array}{ccccc}
0 & 0  & 0       & 0            \\      
0 & 0  & 0       & 0              \\    
0 & 0  & 0       & \beta_4         \\    %
0 & 0  & 0       & 0             \\    %
\end{array} 
\right]
\label{eq:C_W}
\end{equation}
\noindent 
The dynamics of this system are as shown in Eq \ref{eq:dynSingle} using $\mathbf{W}=\mathbf{W}_C$.

\subsection{Stability analysis}
The stability analysis, using non-linear contraction analysis (see Appendix) \cite{LohSlo98,Lohmiller2000,Slotine03,WangSlo}, consists of three steps: i) demonstrate contraction of a single WTA, ii) merge two WTAs by demonstrating that inhibitory units synchronize, and iii) demonstrate contraction of the combined WTAs. We have previously shown how contraction analysis can be applied to reasoning over the stability and functionality of WTA circuits \cite{RutishauserDouglas2010}. Here, we apply and extend the same methods to this new circuit.

Contraction analysis is based on the Jacobians of the system. For purposes of analysis, but without loss of generality, we will base this section on a reduced system with only 1 possible winner for each WTA as shown in Fig \ref{fig:ElegantCircuit}C.

The Jacobian of a single system is
\begin{equation}
\tau \mathbf{J} = \left[ 
\begin{array}{ccc}
\l_1 \alpha - 1  & l_1 -\beta_1 & 0 \\
0                & -1           & l_2 \beta_3 \\
l_3 \beta_2      & 0            & -1 
\end{array} 
\right]
\label{eq:JsingleRed}
\end{equation}
\noindent In a stable network, a constant external input to the first unit $x_1$ will lead to a constant amplitude of $x_1$ that is either an amplified or suppressed version of its input. 

The activation function $f(x)=max(x,0)$ is not continuously differentiable, but it is continuous in both space and time, so that contraction results can still be applied directly \cite{Lohmiller2000}. Furthermore, the activation function is piecewise linear with a derivative of either $0$ or $1$. We exploit this property by inserting dummy terms $l_j$, which can either be $0$ or $1$ according to the derivative of $f(x)$: $l_j=\frac{d}{dx}f( x_j(t) )$. In this case, all units are active and thus $l_1=l_2=l_3=1$.

A system with Jacobian $\mathbf{J}$ is contracting if
\begin{equation}
{\bf \Theta}\ {\bf J} \ {\bf \Theta}^{-1} \ < \ {\bf 0}
\label{eq:TW2}
\end{equation}
\noindent where $\mathbf{\Theta}$ is a constant transformation into an appropriate metric and $\mathbf{F}={\bf \Theta}\ {\bf J} \ {\bf \Theta}^{-1}$ is the generalized Jacobian. If $\mathbf{F}$ is negative definite, the system is said to be contracting. We have shown previous \cite{RutishauserDouglas2010} (section 2.4) how to choose the constant transformation $\mathbf{\Theta}$ and conditions that guarantee contraction for a WTA circuit where all excitatory units provide direct input to the inhibitory unit (Fig \ref{fig:AllCircuits}A). In summary, ${\bf \Theta}={\bf Q}^{-1}$ where ${\bf Q}$ is defined based on the eigendecomposition $\mathbf{J}=\mathbf{Q} \mathbf{\Lambda} \mathbf{Q}^{-1}$. In this case

\begin{eqnarray}
0<\alpha<2 \sqrt{\beta_1 \beta_2 } \nonumber \\
0<\beta_1\beta_2<1
\label{eq:limAlpha}
\label{eq:limBeta1}
\end{eqnarray}

\noindent guarantee contraction for any such WTA of any size \cite{RutishauserDouglas2010}.

Structurally, the two versions of the WTA are equivalent in that an additional unit was added in the pathway of recurrent inhibition, but no inhibition is added or removed (Compare Fig \ref{fig:AllCircuits}A to Fig \ref{fig:ElegantCircuit}A).  Thus, we can apply the same constraints by replacing the product $\beta_1\beta_2$ with $\beta_1\beta_2\beta_3$ in above equations. This product is equivalent to the inhibitory loop gain. This reduction is verified as follows. Using the notation shown in  Fig \ref{fig:ElegantCircuit}C, assume that $T=0$ for $x_3$ where $T>0$ for the other units. Then, 
\begin{eqnarray}
\dot{x}_{3} +  x_{3}  = f( \beta_2 x_1) \nonumber \\
\dot{x}_{2} +  x_{2}  = f( \beta_3 x_3 - T_2)
\end{eqnarray}
\noindent At steady-state, $x_{2}  = f( \beta_2\beta_3 x_1)$, showing that $x_3$ and $x_2$ can be merged into a single unit $x_2$ by providing input of weight $\beta_2\beta_3$ directly to unit $x_2$ (Fig \ref{fig:ElegantCircuit}D). The first key result of this paper are the following limits for contraction of a single such WTA (Fig \ref{fig:ElegantCircuit}A):
\begin{eqnarray}
0<\alpha<2 \sqrt{\beta_1 \beta_2 \beta_3 } \nonumber \\
0<\beta_1\beta_2\beta_3<1
\label{eq:limAlphaNew}
\label{eq:limBeta1New}
\end{eqnarray}

\subsubsection{Synchronizing two WTAs}
Next, we show that connecting two WTAs in the manner illustrated in Fig \ref{fig:ElegantCircuit}C results in synchronization of the two inhibitory units, which in turn leads to the two WTAs merging into a single WTA. Note that by synchronization, we mean that two variables have the same trajectory, or more generally that their difference is constant (in contrast to other meanings of synchronization i.e. in population coding). The approach is to show that adding excitatory connections $\beta_4$ of sufficient strength will lead to the activity of the two inhibitory units $x_2$ and $y_2$ approaching a constant difference.

The Jacobian of the coupled system as shown in Fig \ref{fig:ElegantCircuit}C is
\begin{equation}
\mathbf{J}_C = \left[ 
\begin{array}{cc}
\mathbf{J}_1 & \mathbf{D}_2 \\ 
\mathbf{D}_1 & \mathbf{J}_2
\end{array} 
\right]
\label{eq:Wcoupled2}
\end{equation}

\noindent with $\mathbf{J}_1=\mathbf{J}_2=\mathbf{J}$ (see Eq \ref{eq:JsingleRed}) and

\begin{equation}
\tau \mathbf{D}_{1,2} = \left[ 
\begin{array}{ccccc}
0 &  0       & 0         \\      
0 &  0       & l_{2,5} \beta_4         \\    
0 &  0       & 0         \\    
\end{array} 
\right]
\label{eq:C_W2}
\end{equation}
\noindent Following \cite{PhamSlotine07,RutishauserDouglas2010}, synchronization occurs exponentially if the following holds:
\begin{equation}
\mathbf{V} \mathbf{J}_C \mathbf{V}^T < \mathbf{0}
\label{eq:SyncInhib2}
\end{equation}

\noindent where $\mathbf{V}$ defines an invariant subset of the system such that $\mathbf{V}\mathbf{x}$ is constant, with $\mathbf{x}=(x_1,x_2,x_3,y_1,y_2,y_3)$. Here, we define synchrony as a regime where the differences between the inhibitory units $x_2-y_2$ and between the interconnect units $x_3-y_3$ are constant (although not necessarily zero). This results in 
\begin{equation}
\mathbf{V} = \left[ \begin{array}{cccccc}
0 & 1 & 0               & 0 & -1 &  0\\
0 & 0 & 1               & 0 & 0  & -1\\
\end{array} 
\right]
\label{eq:VsyncTerm2}
\end{equation}
which embeds the two conditions.

Condition (\ref{eq:SyncInhib2}) is satisfied if
\begin{eqnarray}
1<\alpha \nonumber \\
0<\beta_4<\beta_3+2 \\
\beta_3<2
\label{eq:limSync1}
\label{eq:limSync3}
\end{eqnarray}
\noindent The conditions on the interconnect-weight $\beta_4$ guarantees that the dynamics are stable and that the inhibitory units synchronize. As long as $\beta_4>0$ is sufficiently small but non-zero, the inhibitory parts of the system will synchronize. Realistically, $\beta_4$ needs to be sufficiently large to drive the other inhibitory neuron above threshold and will thus be a function of the threshold $T$ (see \cite{RutishauserDouglas2010}, Eq 2.51). Here, synchrony is defined as their difference being constant. This in turn shows that the two WTAs have merged into a single WTA since the definition of a WTA is that each excitatory unit receives an equivalent amount of inhibition (during convergence but not necessarily afterwards, see simulations). This is our second key result.

\subsubsection{Stability of pairwise combinations of WTAs}
The final step of the stability analysis are conditions for the coupling strength $\beta_4>0$ such that the coupled system as shown in Fig \ref{fig:ElegantCircuit}C is contracting. The reasoning in this section assumes that the individual subsystems are contracting (as shown above).

The Jacobian of the combined system remains Eq \ref{eq:Wcoupled2}, where $\mathbf{J}_{1,2}$ are the Jacobians of the individual systems and $\mathbf{C}_{1,2}$ are the coupling terms. Rewriting the Jacobian of the second subsystem $\mathbf{J}_{2}$ by variable permutation $y'_2=y_3$ and $y_3'=y_2$ allows expression of the system in the form of

\begin{equation}
\mathbf{J}_C = \left[ 
\begin{array}{cc}
\mathbf{J}_1 & \mathbf{D_2} \\ 
\mathbf{D}_1 & \mathbf{J}_2
\end{array} 
\right]
=
 \left[ 
\begin{array}{cc}
\mathbf{J}_1 & \mathbf{E} \\ 
\mathbf{E}^T & \mathbf{J}_2
\end{array} 
\right]
\label{eq:WcoupledFlipped}
\end{equation}
\noindent where $\mathbf{E} = \mathbf{D}_1$ (Eq \ref{eq:C_W2}). This transformation\footnote{The variable permutation is equivalent to a transformation of $\mathbf{J}_1$ by the permutation matrix $\bfTheta$: $\mathbf{J}_2=\bfTheta \mathbf{J}_1 \bfTheta^{-1}$
with $\bfTheta = \left[ 
\begin{array}{ccc}
1 & 0 & 0 \\ 
0 & 0 & 1 \\
0 & 1 & 0
\end{array}\right]$.} of $\mathbf{J}_{2}$ is functionally equivalent to the original system (thus, its contraction limits remain) but it allows expression of the connection between the systems in the symmetric form of Eq \ref{eq:WcoupledFlipped}. This functionally equivalent system can now be analyzed using the approach that follows.

A matrix of the form 
$\left[ 
\begin{array}{cc}
\mathbf{J}_1 & \mathbf{D} \\ 
\mathbf{D}^{T} & \mathbf{J}_2
\end{array} \right]$
is negative definite if the individual systems $\mathbf{J}_{1,2}$ are negative definite and if 
$\mathbf{J}_{2}<\mathbf{D}^{T} \mathbf{J}_{1}^{-1} \mathbf{D}$ \cite{Horn85} (Page 472). 
Following \cite{Slotine03} (Section 3.4) and \cite{RutishauserDouglas2010} (Section 2.8), this implies that a sufficient condition for contraction is $\sigma^{2} (\mathbf{D}) < \lambda(\mathbf{J}_1) \lambda(\mathbf{J}_2) $ 
where $\sigma(\mathbf{D})$ is the largest singular value of $\mathbf{D}$ and equivalent to $\beta_4$ in our case (all other elements of $\mathbf{D}$ are zero) and $\lambda$ is the contraction rate of the individual subsystems. Since the two subsystems are equivalent, the contraction rates are also the same $\lambda_1 = \lambda_2 = \lambda(\mathbf{J}_{1,2})$. It thus follows that the coupled systems are stable if $\beta_4 < \lambda_{1}$.

The contraction rate \cite{Slotine03,RutishauserDouglas2010} of an individual subsystem is the absolute value of the largest eigenvalue of the Hermitian part of $\mathbf{F}$ (also see Eq (\ref{eq:TW2})). 
It is $\lambda_{1,2} = | \frac{1}{2}(-2+\alpha)|$ for our system. Thus the condition for the two coupled systems to be contracting reduces to
\begin{equation}
\beta_4 < 1-\frac{\alpha}{2}
\end{equation}

\subsubsection{Summary of boundary conditions}
In summary, the following conditions guarantee stability of both the single and combined system as well as hard competition between the two coupled systems (that is, only one of the WTAs can have a winner). These conditions can be relaxed if $\alpha<1$, which will permit a soft winner-take-all.

\begin{eqnarray}
1 < \alpha < 2 \sqrt{\beta_1\beta_2\beta_3} \nonumber \\
0 < \beta_1\beta_2\beta_3 < 1 \nonumber \\	
0 < \beta_4 < 1-\frac{\alpha}{2}
\label{eq:allBounds}
\end{eqnarray}
\noindent The lower bound on $\beta_4$ is from the synchronization analysis, whereas the upper bound is from the stability analysis. These results illustrate the critical tradeoff between having enough strength to ensure functionality, while being weak enough to exclude instability.

\subsection{Speed of winner selection}
How quickly will a system select a winner? For a single WTA, this question is answered by how quickly a single system contracts toward a winner and for a coupled system how quickly the two systems synchronize. One of the key advantages of contraction analysis is that the rate of contraction, and in this case the rate of synchronization, can be calculated explicitly. We will express the contraction and synchronization rate in terms of the time constants $\tau$ and its inverse, the decay constant. $\tau$ refers to the mean lifetime of exponential decay $x(t)=N_0 e^{-t\frac{1}{\tau} } = N_0 e^{-\lambda t}$. $\lambda=\frac{1}{\tau}$ is the decay constant. Both the contraction and the synchronization rate are expressed in the form of a decay constant $\lambda$. For example, the contraction rate of a system of the form $\dot{N}=-\lambda N$ is equivalent to $\lambda$. 

Physiologically, the time constants $\tau$ in our system are experimentally determined membrane time constants that are typically in the range of 10-50ms \cite{Koch1998,McCormickEtal1985,Koch96,Brown81}. For simplicity, we assume that all excitatory and inhibitory units have the same, but different, time constants $\tau_E$ and $\tau_I$, respectively. While the exact values depend on the cell type and state of the neurons, it is generally the case that $\tau_I<\tau_E$ due to the smaller cell bodies of inhibitory neurons \cite{McCormickEtal1985,Koch1998}. 

The bounds (\ref{eq:allBounds}) were calculated assuming equal time constants for all units. However, the same calculations yield to very similar bounds when assuming different time constants for inhibitory and excitatory units (Appendix C, \cite{RutishauserDouglas2010}). In this case the ratio of the time constants $\frac{\tau_I}{\tau_E}$ becomes an additional parameter for the parameter bounds.

\subsubsection{Speed of synchronization}

The synchronization rate is equivalent to the contraction rate of the system defined in Eq \ref{eq:SyncInhib2} \cite{PhamSlotine07}, which is the absolute value of the maximal eigenvalue of the Hermitian part of  $ \mathbf{V'} (\mathbf{R} \mathbf{J}_C) \mathbf{V'}^T$. Here, the original $\tau^{-1}$ is replaced by the diagonal matrix $\mathbf{R}$, with the appropriate $\tau_E,\tau_I$ terms on the diagonal \footnote{For the example of $\mathbf{J}_C$ (Eq \ref{eq:Wcoupled2}), the diagonal terms are $\tau_E^{-1},\tau_E^{-1},\tau_I^{-1},\tau_E^{-1},\tau_E^{-1},\tau_I^{-1}$}. The matrix $\mathbf{V'}$ is an orthonormal version of $\mathbf{V}$ as defined in Eq \ref{eq:VsyncTerm2}, which here is simply $\mathbf{V'}=\sqrt{2^{-1}}\mathbf{V}$.

The resulting synchronization rate (sync rate) is a function of the weights $\beta_3$ (local inhibitory loop) and $\beta_4$ (remote inhibitory loop). We assume $\beta_4 \le \beta_3$, which means that remote connectivity is weaker than local connectivity. However, qualitatively similar results can be found using the opposite assumption. For $\tau_E=\tau_I$, the sync rate is  

\begin{equation}
\lambda_s = \frac{1}{2 \tau}(2-\beta_3+\beta_4)
\label{eq:syncRate1}
\end{equation}

\noindent Note the tradeoff between local and remote connectivity: stronger remote connectivity increases and stronger local connectivity decreases the speed of synchronization (the larger $\lambda$, the quicker the system synchronizes). For approximately equal connectivity strength $\beta_3 \simeq \beta_4$ or in general for $\beta_3,\beta_4 \ll 1$, the sync rate is approximately $\tau^{-1}$.

In general for $\tau_E \neq \tau_I$, the sync rate is $\lambda_s = \frac{\tau_E + \tau_I-\sqrt{\tau_E^2-2 \tau_E \tau_I+\left(1+(\beta_3-\beta_4)^2\right) \tau_I^2}}{2 \tau_E \tau_I}$. For example, for $\tau_I=\frac{\tau_E}{2}$ this reduces to $\lambda = \frac{1}{2 \tau_I}(3-\sqrt{1+4(\beta_3-\beta_4)^2})$. Again, for $\beta_3 \simeq \beta_4$, the sync rate is approximately $\tau_I^{-1}$. In conclusion, the sync rate is thus approximately equal to the contraction rate of the inhibitory units. Thus, synchronization occurs very quickly (20-50ms for typical membrane time constants).

\subsubsection{Speed of contraction}
The speed of selecting a winner (the contraction rate) for a single WTA can similarly be calculated based on the absolute value of the maximal eigenvalue of the Hermitian part of ${\bf \Theta} ({\bf R}{\bf J}) {\bf \Theta}^{-1}$ (\ref{eq:TW2}).

Assuming $\tau=\tau_E=\tau_I$, the contraction rate is

\begin{equation}
\lambda_c = \frac{1}{2 \tau}(2-\alpha)
\label{eq:syncRate2}
\end{equation}

\noindent Note that the larger $\alpha$, the longer it takes till the system converges. Qualitatively similar findings result for other ratios of $\tau_E$ and $\tau_I$. For a typical value of $\alpha=1.2$ (see simulations below) and $\tau=20ms$, the contraction rate would be $20s^{-1}$. This equals a half-way time (time constant) of $\lambda^{-1}=50ms$. For $\alpha=1.5$, this would increase to $80ms$. The time it takes to find a winner is thus a multiple of the membrane time constant (in this example $20ms$) and substantially slower than the time it takes to synchronize the network. In conclusion, synchronization is achieved first which is then followed by winner selection.

\subsection{Coupling more than two WTAs}

So far we have shown how two different WTAs compete with each other after their inhibitory neurons are coupled. Similarly, more than two WTAs can compete with each other by all-to-all coupling of the inhibitory units, i.e. every WTA is connected with two $\beta_4$ connections from and to every other WTA. Thus, the wiring complexity of this system scales as $O(M^2)$ where $M$ is the number of WTAs in the system (note that $M$ is not the number of units but the number of WTAs). Notice also that the all-to-all coupling concerns only the sparse long-range excitatory connections and not the internal connectivity of the WTAs them-self.

The same principle can be used to embed hierarchies or sequences of competition. Thus, in a network of such WTAs, some WTAs could be in direct competition with each other while others are not. Thus, for example, in a network of three WTAs A, B, and C relationships such as A competes with B and B competes with C are possible. In this case A does not directly compete with C. So if A has a winner, C can also have a winner. If B has a winner, however, neither B nor C can have a winner (see Fig \ref{fig:Sim2}D-F for a demonstration).

Regardless of how many WTAs are combined and whether all compete with all or more selectively, the stability of the aggregated system is guaranteed if the individual sub-systems are stable and the coupling strengths $\beta_4$ observe the derived bounds. While in themselves combinations of stable modules have no reason to be stable, certain combinations (such as the one we utilize) of contracting systems are guaranteed to be stable \cite{Slotine01}. This is a key benefit of Contraction Analysis for the analysis of neural circuits.

\subsection{Numerical simulations}

\begin{figure}
\centering
\includegraphics[angle=0,width=15cm]{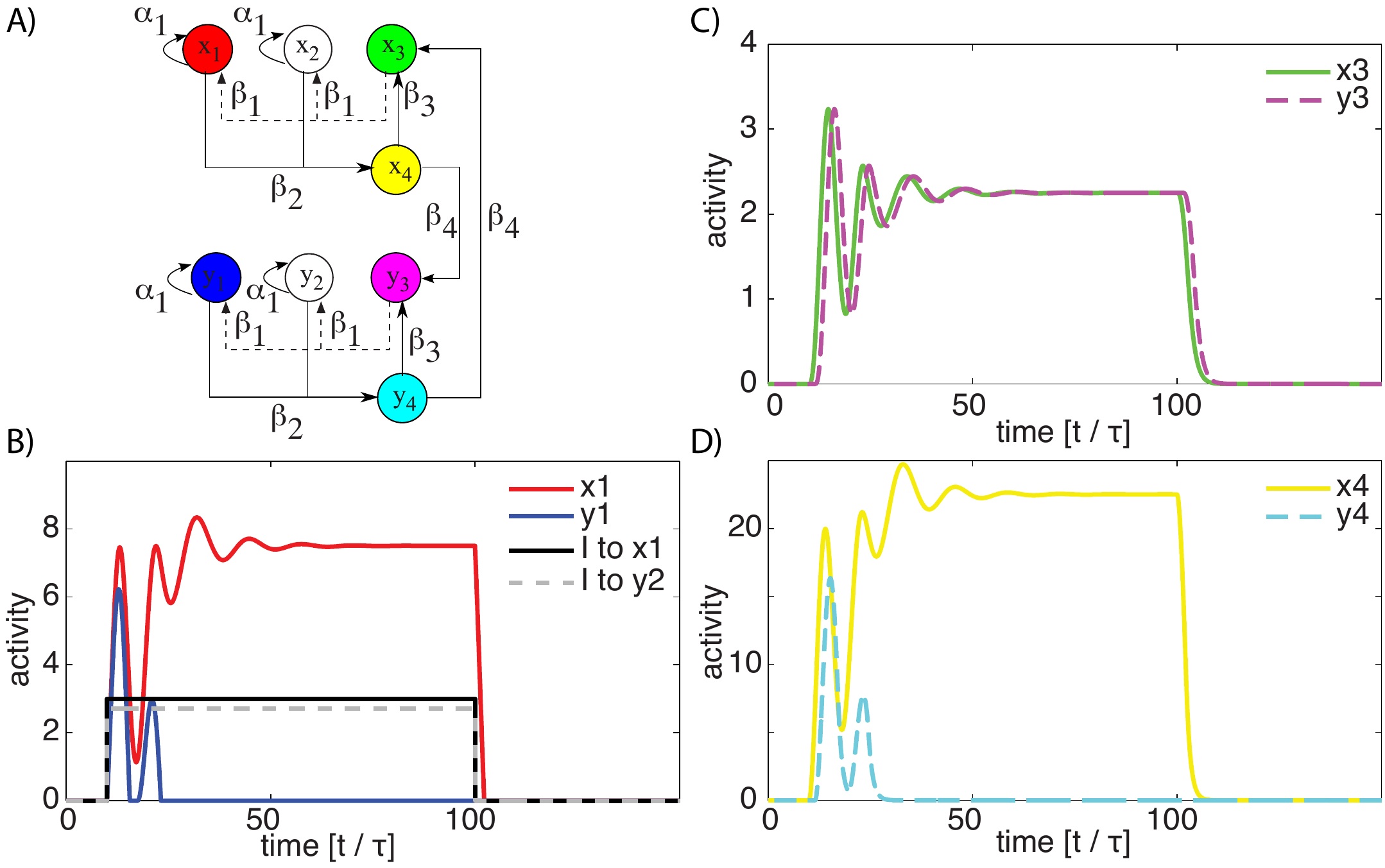}
\caption{Simulation of merged WTA consisting of two WTAs with two excitatory units each (four possible winners). 
(A) Illustration of connectivity and notation (color code).
(B) Activity as a function of time, for two excitatory units on two different WTAs, together with the external input provided to the same units. Notice how the network selects the winner appropriately.
(C) Activity of the inhibitory units in each WTA. Note: $y_3$ is slightly delayed for plotting purposes. 
(D) Activity of the two interconnect units. Notice how the output of the losing WTA $y_4$ descends to zero after the competition has been resolved and the network has contracted.
Units of time are multiplies of the time constant $\tau$. Notice that the same color indicates the same unit throughout this figure (notation is shown in A).
}
\label{fig:Sim1}
\end{figure}

We simulated several cases of the network to illustrate its qualitative behavior. We used Euler integration with $\delta=0.01s$. The analytically derived bounds offer a wide range of parameters for which stability as well as function is guaranteed. For the simulations, we chose parameters that verify all bounds discussed.

First, we explored a simple system consisting of two WTAs with two possible winners each (Fig \ref{fig:Sim1}). Parameters were $\alpha=1.2,\beta_1=2,\beta_2=3,\beta_3=\beta_4=0.1$ and $T=0$. We found that any of the four possible winners can compete with each other irrespective of whether they reside on the first or second WTA (Fig \ref{fig:Sim1}B-D shows an example). The inhibitory units quickly synchronized (Fig \ref{fig:Sim1}C) their activity and reached the same steady-state amplitude (because $\beta_3=\beta_4$)\footnote{If $\beta_3=\beta_4$, it can be verified that $x_3(t)=y_4(t)$ for all $t>0$ if the initial values at $t=0$ are equal. Thus, the two inhibitory neurons become exactly equivalent in this special case.}. 

\begin{figure}
\centering
\includegraphics[angle=0,width=15cm]{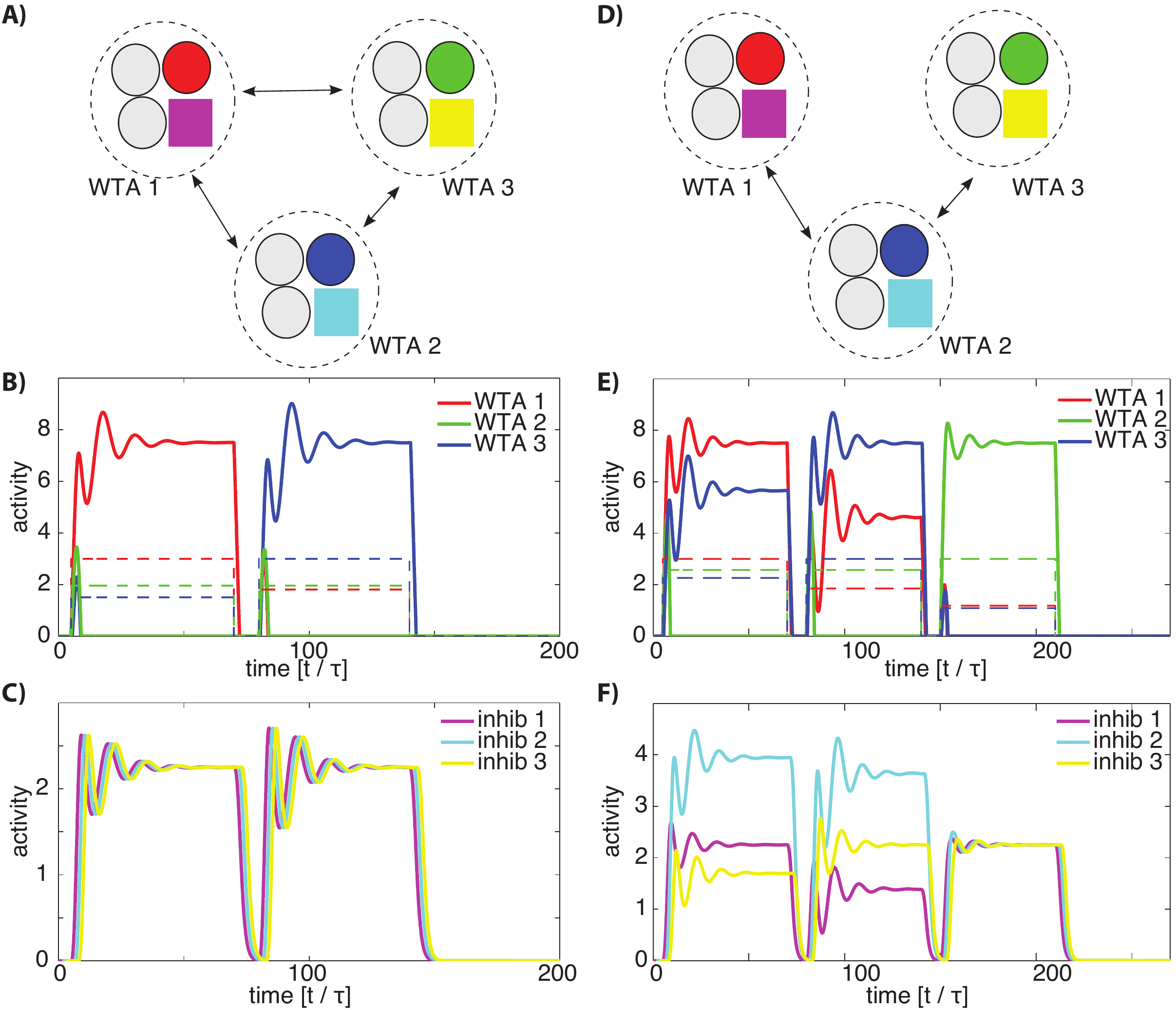}
\caption{Simulation of merged WTA consisting of three WTAs with three possible winners each.
(A-C) Case 1: Pairwise coupling allows all-to-all competition. 
(A) Illustration of connectivity. Filled circles are excitatory neurons, filled rectangles inhibitory. The activity of
units with colored fills are shown as a function of time.
(B) Activity of one excitatory unit for each WTA (bold lines) and the external input to each (dashed lines). Notice that there can be only one winner among the three WTAs. 
(C) Activity of the inhibitory units, shifted in time to each other slightly for plotting only.
(D-F) Case 2: selective coupling, allowing partial competition only between 1\&2 and 2\&3 but not 1\&3.
(E) Activity of the excitatory units for different cases. Notice that WTA 1 and 3 can have winners simultaneously, but not 2.
Numbers indicate the WTA the winner belongs to.
(F) Activity of the inhibitory units, illustrating synchrony in the presence of different absolute amplitudes.
Units of time are multiplies of the time constant $\tau$. Notice that the same color indicates the same unit throughout this figure (notation is shown in A).
} 
\label{fig:Sim2}
\end{figure}

Second, we simulated a system with 3 WTAs using the same parameters (Fig \ref{fig:Sim2}). For all-to-all coupling, all 3 WTAs directly compete with each other (Fig \ref{fig:Sim2}A,B), i.e. there can only be one winner across the entire system. Again, the inhibitory units all synchronize quickly during and after convergence (Fig \ref{fig:Sim2}C). We also simulated the same system with more selective connectivity, eliminating competition between WTAs 1 and 3 (Fig \ref{fig:Sim2}D). This arrangement allows either one winner if it is on WTA 2, or two winners if they are on WTAs 1 and 3. If the maximal activity is not on WTA 2, then the network permits 2 winning states. Otherwise, if the maximal input is on WTA 2 only 1 winner is permitted  (see Fig \ref{fig:Sim2}E for an illustration). This configuration allows for partial competition.

\section{Discussion}

Neural circuits commonly depend on negative feedback loops. Such recurrent inhibition is a crucial element of microcircuits from a wide range of species and brain structures \cite{shepherd2010handbook} and enables populations of neurons to compute non-linear operations such as competition, decision making, gain control, filtering, and normalization.  However, when considering biologically realistic versions of such circuits additional factors such as wiring length, specificity and complexity become pertinent. Here, we are principally concerned with the superficial layers of neocortex where the average distance of intracortical inhibitory connections is typically much shorter than the excitatory connections \cite{BockReid11,Binzegger04,PerinMarkram11,KatzelMiesenbock11,AdesnikScanziani10}.
In contrast, in invertebrates an inhibitory neuron has been identified that receives input from and projects back to all Kenyon cells (which are excitatory) \cite{Papadopoulou11}. This neuron has been demonstrated to perform response normalization, making this system a direct experimental demonstration of competition through shared inhibition. No such system has yet been identified in the cortex.

The number of excitatory neurons that can be contacted by an inhibitory neuron thus poses a limit on how many excitatory neurons can compete directly with one another (in terms of numbers and distance). Other models, such as those based on Mexican-hat type inhibitory surrounds \cite{Hertz91,WillshawMalsburg76,SoltaniKoch2010}, even require that inhibitory connectivity be longer range than the excitatory. These anatomical constraints have been used to argue that models such as the WTA are biologically unrealistic and as such of limited use.

We have demonstrated here, by theoretical analysis and simulation, that it is possible to extend such circuits by merging several independent circuits functionally, through synchronization of their inhibitory interneurons. This extension allows the construction of large, spatially distributed circuits that are composed of small pools of excitatory units that share an inhibitory neuron. We have applied and proved by non-linear contraction analysis that systems combined in this manner are inherently stable and that arbitrary aggregation by inhibitory synchrony of such sub-systems results in a stable system. This composition of subcircuits removes the limits on maximal circuit size imposed by anatomical wiring constraints on inhibitory connectivity, because the synchrony between local inhibitory neurons is achieved entirely by excitatory connectivity which can possibly be long-range so permitting competition between excitatory units that are separated by long distances; for example, in different cortical areas. We show that the time necessary to achieve sychronization is much shorter than the time required to select a winner. Thus, synchronization is faster than winner selection, which can thus proceed robustly across long-range connections that enforce synchronization. 
Further, selective synchronization between some WTAs but not others allows partial competition between some but not other WTAs (see Fig \ref{fig:Sim2}). The strength of these long-range connections could be modulated dynamically to enable/disable various competitions between two populations conditional on some other brain state. This modulation could be implemented by a state-dependent routing mechanism \cite{RutishauserDouglas2009}.

There are several possibilities of mapping the abstract units in our model to real physiological neurons. Our units are mean-rate approximations of a small group of neurons. In terms of intra-cortical inhibition, these would lie anatomically close to each other within superficial layers of neocortex. Since such inhibitory connectivity would have only limited reach, each inhibitory subunit can only enforce competition across a limited number of closeby excitatory units. 
Competition between different areas is made possible by synchronizing remote populations by long-range excitatory mechanisms in the way we propose. 
Direct long-range inhibition, on the other hand, is unlikely both intracortically and subcortically, since all known connections from the thalamus and basal ganglia to cortex are excitatory. Networks such as the LEGION network \cite{WangTerman95} assume global inhibitory input to all excitatory units in the network, which for the reasons we discuss is unlikely in the case of cortex. It would, however, be possible to implement a feasible version of the global inhibitory input by synchronizing many local inhibitory neurons using the mechanism we describe, resulting in an anatomically realistic version of the LEGION network.

Functionally, the model presented here makes several testable predictions. Consider a sensory area with clearly defined features as possible winners, such as orientations. The model predicts that the inhibitory units would not be tuned to these features, particularly if the number of possible winners is large. This is because the connectivity to the inhibitory units is not feature specific. Experimental studies indicate that this is indeed the case: units that functionally represent different tuning project to the same inhibitory unit, resulting in untuned inhibitory activity \cite{BockReid11,FinoYuste11,Kerlin10,Kuhlman11,Hofer11}.
Secondly, this model predicts that inhibitory activity between two different areas or parts of the same area can either be highly synchronous or completely decoupled depending on whether at present the two are competing or functioning independently. This thus predicts that synchrony of inhibitory units should be affected by manipulations that manipulate competition, such as top-down attention.

Our model suggests that synchronized populations of inhibitory neurons are crucial for enforcing competition across several subpopulations of excitatory neurons. It further suggests that the larger the number and spatial distribution of such synchronized inhibitory units, the larger the number of units that compete with each other. Experimentally, synchronized modulation of inhibitory neurons is a common phenomena that is believed to generate the prominent gamma rhythm triggered by sensory stimulation in many areas \cite{Fries07,Whittington95,Traub96}. Recent experiments have utilized stimulation of inhibitory neurons \cite{Cardin09,Sohal09,Szucs09} to increase or decrease their synchronization with direct observable effects on nearby excitatory neurons such as, for example, increased or decreased amplitude and precision of evoked responses relative to how strongly the inhibitory neurons were synchronizing. Note that our proposal for this function of inhibitory synchrony is distinct and independent from the proposal that gamma-band synchrony serves to increase readout efficacy by making spikes arrive co-incidentally from a large number of distributed sources \cite{Tiesinga08,SingerGray95}. Here, we propose that an additional function of such synchrony is to allow select populations of excitatory neurons to compete with each other because they each receive inhibition at the same time.

\section{Appendix: Contraction Analysis}
This section provides a short summary of contraction analysis. We have previously published the detailed methods of applying contraction theory to WTA circuits \cite{RutishauserDouglas2010}.
Essentially, a nonlinear time-varying dynamic system will be called {\it contracting} if arbitrary initial conditions or temporary disturbances are forgotten exponentially fast, i.e., if trajectories of the perturbed system return to their unperturbed behavior with an exponential convergence rate. A relatively simple algebraic conditions can be given for this stability-like property to be verified, and this property is preserved through basic system combinations and aggregations.

A nonlinear contracting system has the following properties~\cite{LohSlo98,Lohmiller2000,Slotine03,WangSlo}
\begin{itemize}
	\item global exponential convergence and stability are guaranteed
	\item convergence rates can be explicitly computed as eigenvalues of well-defined Hermitian matrices
	\item combinations and aggregations of contracting systems are also contracting
	\item robustness to variations in dynamics can be easily quantified
\end{itemize}

Before stating the main contraction theorem, recall first the following properties: The symmetric part of a matrix $\mathbf{A}$ is $\mathbf{A}_H=\frac{1}{2}(\mathbf{A}+\mathbf{A}^{*T})$. A complex square matrix $\mathbf{A}$ is {\it Hermitian} if $\bfA^T = \bfA^*$ , where $^T$ denotes matrix transposition and $^*$ complex conjugation. The {\it Hermitian part} $\bfA_H$ of any complex square matrix $\bfA$ is the Hermitian matrix $\frac{1}{2}(\bfA + \bfA^{*T})$ . All eigenvalues of a Hermitian matrix are {\it real} numbers.  A Hermitian matrix ${\bf A}$ is said to be {\it positive definite} if all its eigenvalues are strictly positive. This condition implies in turn that for any non-zero real or complex vector ${\bf x}$, ${\bf x}^{*T}{\bf A}{\bf x} > 0$. A Hermitian matrix ${\bf A}$ is called {\it negative definite} if $ - {\bf A}$ is positive definite.

A Hermitian matrix $\mathbf{A}(\mathbf{x},t)$ dependent on state or time will be called {\it uniformly} positive definite if there exists a strictly positive constant such that for all states $\mathbf{x}$ and all $t \ge 0$ the eigenvalues of  $\mathbf{A}(\mathbf{x},t)$ remain larger than that constant. A similar definition holds for uniform negative definiteness.

Consider now a general dynamical system in $\mathbb{R}^n$,
\begin{equation}
\label{eq:main}
	\dot \bfx = \bff(\bfx,t)
\end{equation}
with $\bff$ a smooth non-linear function. The central result of Contraction Analysis, derived in~\cite{LohSlo98} in both real and complex forms, can be stated as:

{\bf Theorem}
	Denote by $\frac{\partial \bff} {\partial \bfx}$ the Jacobian
        matrix of $\bff$ with respect to $\bfx$.  Assume that there exists a
        complex square matrix $\bfTheta(x,t)$ such that the Hermitian
        matrix $\bfTheta(x,t)^{*T}\bfTheta(x,t)$ is uniformly positive
        definite, and the Hermitian part ${\bf F}_H$ of the matrix 
	\[
        \bfF = \left(\dot\bfTheta + \bfTheta \frac{\partial \bff} {\partial \bfx}
        \right) \bfTheta^{-1} 
	\] 
        is uniformly negative definite. Then, all system trajectories converge
        exponentially to a single trajectory, with convergence rate
        $|\sup_{\bfx,t}\lambda_\mathrm{max}({\bf F}_H)|>0$. The system is said to
        be \emph{contracting}, $\bfF$ is called its \emph{generalized
        Jacobian}, and $\bfTheta(x,t)^{*T}\bfTheta(x,t)$ its contraction
        \emph{metric}. The contraction rate is the absolute value of the largest eigenvalue (closest to zero, although still negative) $\lambda = | \lambda_{max}\mathbf{F}_H |$. 

In the linear time-invariant case, a system is globally contracting if and only if it is strictly stable, and $\bfF$ can be chosen as a normal Jordan form of the system, with $\bfTheta$ a real matrix defining the coordinate transformation to that form~\cite{LohSlo98}. Alternatively, if the system is diagonalizable, $\bfF$ can be chosen as the diagonal form of the system, with $\bfTheta$ a complex matrix diagonalizing the system. In that case, $\bfF_H$ is a diagonal matrix composed of the real parts of the eigenvalues of the original system matrix.

\end{document}